\documentclass{elsart}
\usepackage{epsfig,amssymb}
\usepackage{rotating}

\begin{document}

\begin{frontmatter}

% Title, authors and addresses

% use the thanksref command within \title, \author or \address for footnotes;
% use the corauthref command within \author for corresponding author footnotes;
% use the ead command for the email address,
% and the form \ead[url] for the home page:
% \title{Title\thanksref{label1}}
% \thanks[label1]{}
% \author{Name\corauthref{cor1}\thanksref{label2}}
% \ead{email address}
% \ead[url]{home page}
% \thanks[label2]{}
% \corauth[cor1]{}
% \address{Address\thanksref{label3}}
% \thanks[label3]{}

\title{Intense Source of Slow Positrons}

% use optional labels to link authors explicitly to addresses:
% \author[label1,label2]{}
% \address[label1]{}
% \address[label2]{}

\author{P.Perez} 
\ead{patrice.perez@cea.fr} 
\author{A. Rosowsky}
\ead{andre.rosowsky@cern.ch}
\address{DSM/Dapnia/SPP, CEA/Saclay, F-91191 GIF-SUR-YVETTE}

\begin{abstract}
We describe a novel design for an intense source of slow positrons based 
on pair production with a beam of electrons from a 10 MeV accelerator 
hitting a thin target at a low incidence angle. The positrons are collected
with a set of coils adapted to the large production angle. 
The collection system is designed to inject the positrons in a
Greaves-Surko trap~\cite{traps}.
Such a source could be the basis for a series of experiments in fundamental 
and applied research and would also be a prototype source for industrial 
applications which concern the field of defect characterization in the 
nanometer scale.
\end{abstract}

\begin{keyword}
% keywords here, in the form: keyword \sep keyword
positron \sep
positronium \sep
accelerator/linac \sep
defect characterization \sep
3D molecule imaging \sep
energy storage \sep
gamma ray laser \sep
matter antimatter symmetry \sep
antigravity

% PACS codes here, in the form: \PACS code \sep code
\PACS code \sep code
\end{keyword}
\end{frontmatter}

\section{Introduction}
Phenomena involving positrons are important in many fields of physics, 
including astrophysics, plasma physics, atomic physics and materials science.
In the laboratory, low energy positrons are now being used for many of these 
applications, including study of electron-positron plasma 
phenomena~\cite{plasma2}, atomic and molecular physics~\cite{atom2}, 
antihydrogen formation~\cite{exp-ad}, modeling of astrophysical 
processes~\cite{astro2}, and the characterization of 
materials~\cite{materials}.
The limitations in these studies are often due to the relative inavailability 
of suitable positron sources.  

Methods to accumulate, cool and manipulate 
positron plasmas have been developed~\cite{traps}
%in particular in the group of Prof. 
%C. Surko at U.C. San Diego (ref).  
%A specially designed Penning-Malmberg trap using inelastic collisions with
%an N2 buffer gas provides the most efficient antimatter accumulation technique
%to date (refs). It is produced for sale commercially 
and used in a recent experiment that created low energy antihydrogen atoms at 
CERN~\cite{exp-ad}. 
Long confinement times (days), increased plasma densities and 
brightness enhancement are achieved with the rotating electric field 
technique~\cite{rotatfield}
at temperatures of order 1 meV ($\approx$ 10 K). 
%The near future objectives are to store more than $10^{10}$ positrons 
%for days at temperatures of order 1 meV ($\approx$ 10 K).\\

The apparatus presented in this article is designed to produce fluxes of slow
positrons of order $10^{10}$ per second and be stored and/or cooled in a
trap.
% of the above mentioned Greaves-Surko design. 
In contrast, radioisotope 
sources currently in use produce fluxes $\leq 2 \ 10^7 \mathrm{s^{-1}}$.
Its size is also much smaller than that of a large linac or of a nuclear 
reactor but could be used as a small facility for 
interdisciplinary experimental studies with positrons.

Such a source would allow the production of positronium (Ps), hydrogen and 
antihydrogen in a symmetric way in the same experimental conditions through 
the set of reactions: 
$e^+ + e^- + e^{\pm}\rightarrow Ps + e^{\pm} $, followed by
$Ps + p \rightarrow H + e^+$ or its antimatter counterpart 
$Ps + \overline{p} \rightarrow \overline{H} + e^-$ and the production of ions
via $H + Ps \rightarrow H^- + e^+ $ or 
$\overline{H} + Ps \rightarrow \overline{H}^+ + e^- $~\cite{our-antiH}.

A novel method to produce a 3D image of molecules with a resolution of few 
\AA ngstr\"{o}ms has been proposed~\cite{mills-imaging}. 

Since several years the possibility to create a Bose-Einstein Condensate 
(BEC) of positronium and to make a 511 KeV gamma ray 
laser is being studied~\cite{mills-laser}~\cite{Liang}. 

These techniques relie or would benefit a lot on 
the availability of a very intense source of positrons. 
Moreover a high 
production rate of slow positrons (exceeding $10^{10} \rm{s^{-1}}$) and of
positronium is being looked for in industrial and research applications, 
for instance, ``Positron Annihilation Spectroscopy'' (PAS)~\cite{PALS}.

The most commonly used source of positrons is $\rm{^{22}Na}$. Such compact 
sources are well suited for laboratory research, but their maximum activity
lies around $4 \ 10^9$ Bq with a mean lifetime of 2.6 years. There are also
some accelerators (100 MeV) partly used for the production of slow positrons 
which are managed as a facility and a nuclear research reactor in 
Munich~\cite{frm2}.

We propose an alternative intense 
($>10^{12} \rm{s^{-1}}$) source of slow (MeV) positrons based on $e^{+}e^{-}$
pair creation through the interaction of a 10 MeV electron beam on a target
with an intensity of a few mA.
This source was designed to be coupled with a Greaves-Surko 
trap~\cite{surko-trap} in order to produce a bright beam of slow positrons 
(meV to KeV). 
It may also be used to produce positronium by applying the beam onto a 
cristal~\cite{mills-alsio2}.

The pair production cross section increases with energy. However the 
development of such a setup for university or industrial applications 
limits the beam energy to 10 MeV because a higher energy would start to 
activate the environment (legal limit). 
A dedicated facility would have a size comparable to that of a radioactive 
source but requires a special building for shielding.

\section{Production of positrons}
The positrons are produced by the interaction of a flat electron beam
with a 50 microns target foil. The
angle between the beam plane and the foil is very small,
approximatively 3 degrees. The positron kinetic energy spectrum is
peaked at 1.2 MeV and extends to 8 MeV.

The first step in the positron capture by the trap is the moderation
process. This process involves the slowing down of the positrons, the
creation of meta-stable states with collective charge oscillations in
the moderator and its re-emission at  $\approx$ 1.5 eV. The moderation
efficiency decreases with the incident positron kinetic energy and is
negligible at a few MeV. Therefore a magnetic collector was designed to
separate the positrons from the electrons while preserving the positrons
with a kinetic energy below 1 MeV.

\begin{figure}
%\vspace{25mm}
\begin{center}
\begin{picture}(150,220)
\put(-140,0){\epsfxsize70mm\epsfbox{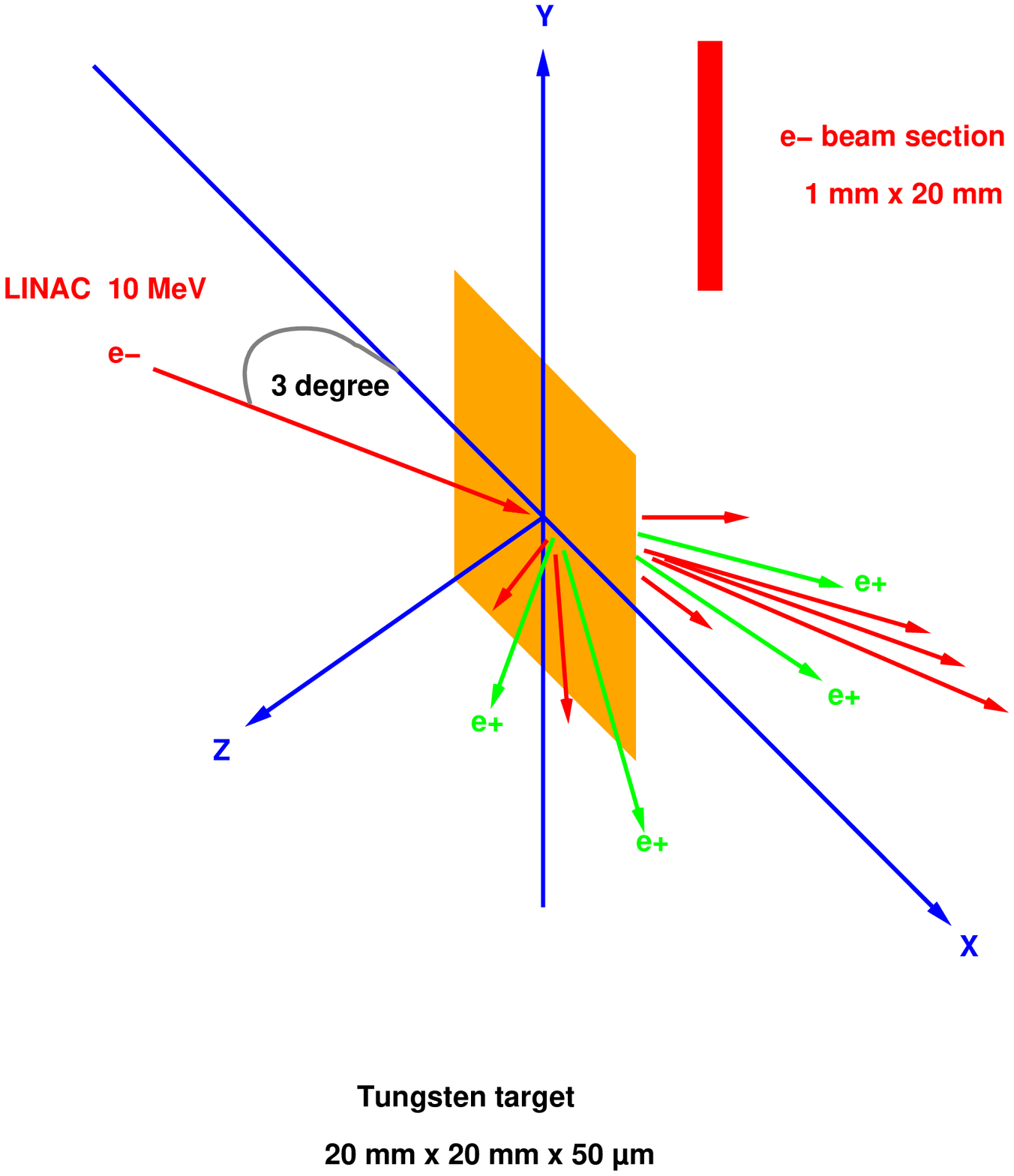}}
\put(80,60){\epsfxsize70mm\epsfbox{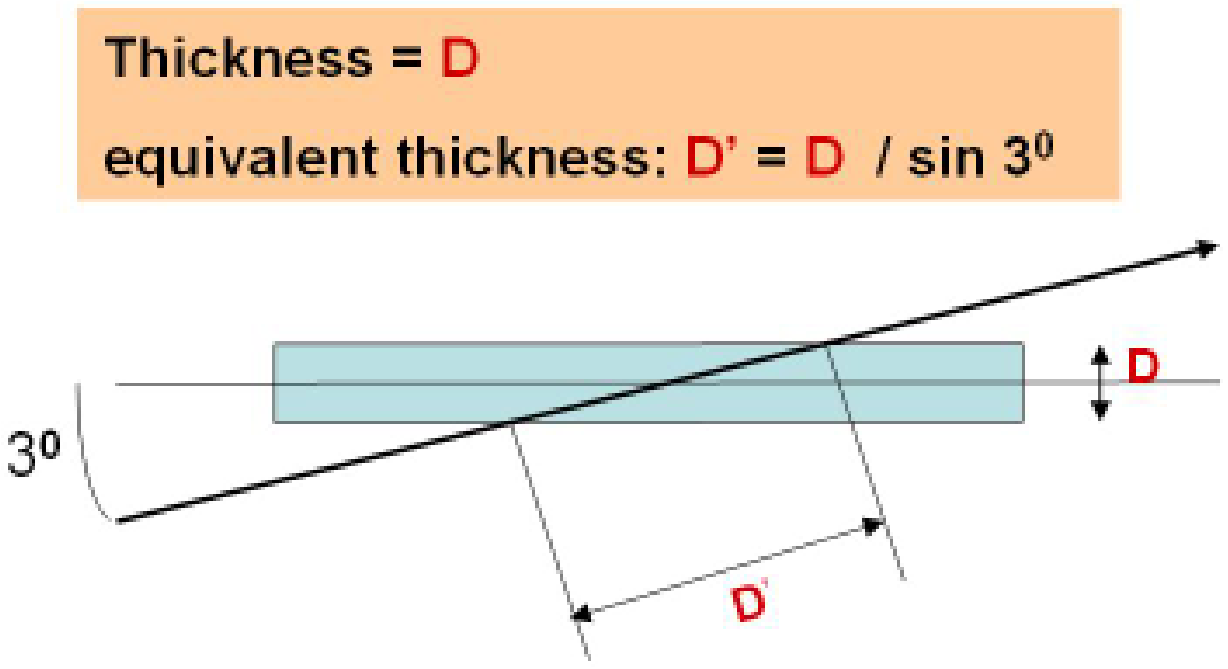}}
\end{picture}
\end{center}
\vspace*{-0.25cm} 
\caption{\protect\footnotesize Diagram of the simulated target (left). 
Sketch showing the equivalent thickness definition (right).}
\label{fig:simul-cible}
\end{figure}

\subsection{Target}
The positron rate is limited by the heating of the target.
The target material is tungsten because of its 
high fusion point (3695K). 

\begin{figure}
%\vspace{-1cm}
\begin{center}
%\begin{picture}(100,90)
%\put(10,10){\epsfxsize85mm\epsfbox{fig-soudure1.eps}}
\includegraphics*[width=65mm]{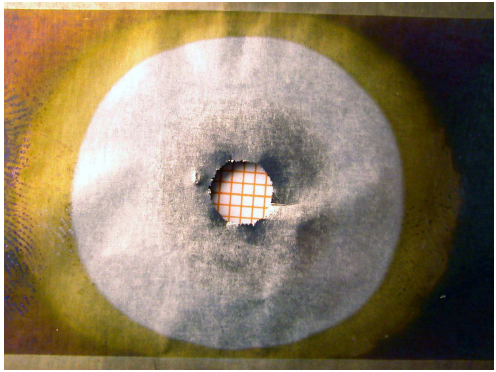}
%\end{picture}
\end{center}
\vspace*{-0.3cm} 
\caption{\protect\footnotesize Perforation of a 50 $\mu$m tungsten sample under
the electron soldering gun. The scale is given by the millimeter paper 
underneath.}
\label{fig:soudure1}
\end{figure}

An experimental test was performed with an electron gun used to weld metal
pieces. The gun delivered a beam with a diameter around 5mm, 
i.e. a surface of around 20 $\rm{mm^2}$~(fig~\ref{fig:soudure1}). 
The 99.99\% purity tungsten target we 
tested had a thickness of 50 $\mu$m~\cite{goodfellow}, dimensions of 
5 cm x 5 cm and was held with a piece of tungsten surrounding it on three of 
its sides. 
The accelerating voltage of the gun was fixed at 40 kV, the intensity was 
gradually increased until perforation at 20 mA. 
%Stefan's law predicts a temperature of 3690K, thus compatible with fusion.
A 15 mA 
current does not perforate the target. At 40 kV electrons deposit all their
energy in the metal. The target sustains thus a deposit greater than 
2 kW/$\rm{cm^2}$. However a 1 kW/$\rm{cm^2}$ limit is set on the 
deposited power as a safety margin.

Even if the temperature is kept well below the fusion point, there will be 
metal evaporation. Using tables from Langmuir and Jones~\cite{langmuir},
on the evaporation rate of tungsten filaments for light bulbs under the 
Joule effect, the target would
lose 10\% of its mass in one hour at 3100 K, and 24 hours at 2700 K. 
A simple way of operation would then consist in exchanging the tungsten foil
every night. 
%This is also adequate with
%the running conditions with a Greaves-Surko trap, which needs to regenerate 
%the solid neon moderator every 24 h.
Running at 2700 K means a lower electron intensity and thus a reduction in the 
$e^+$ rate of a factor 1.7. 
%We foresee a test of evaporation in conditions very similar to this project:
%we have agreed with the IBA~\cite{IBA} firm to put a target sample in one of 
%their intense 10 MeV electron beam lines and measure the temperature rise as 
%well as the mass loss and evaporation depth profile for foils of different 
%thicknesses.\\

%\subsection{Simulations}
%The aim of this section is to optimize the rate of positrons with less than 
%1 MeV kinetic energy.

\subsection{Energy deposit in the target}

Simulations were performed with the GEANT 3.21~\cite{geant} fortran library.
Electrons of 10 MeV/c momentum were generated at an incidence angle of 3 or 90 
degrees w.r.t the target plane. For the 3 degree incidence angle, the 
transverse shape of the beam is a rectangle of 1 mm x 20 mm thus illuminating 
a square area of 20 mm x 20 mm on the target.

\begin{figure}
%\vspace{1.cm}
\begin{center}
\begin{picture}(150,220)
\put(-140,0){\epsfxsize80mm\epsfbox{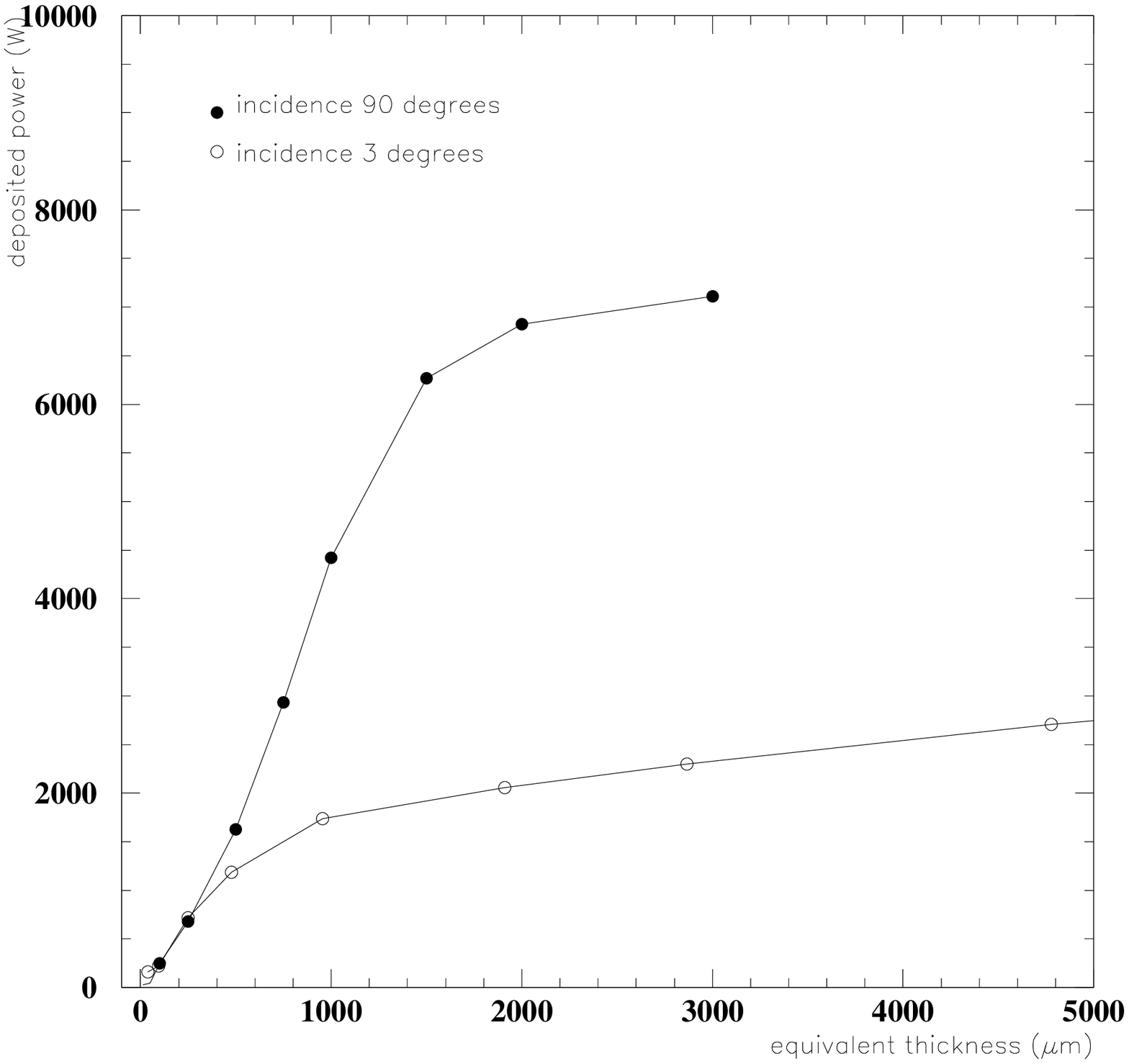}}
%\includegraphics*[width=80mm]{heat-10.eps}
%\end{picture}
%\end{center}
%\vspace*{-2cm} 
%\caption{\protect\footnotesize Power deposit as a function of the equivalent 
%thickness crossed for 3 and 90 degree incidence angles and a beam of 10 MeV 
%energy and 1 mA current.}
%\label{fig:depot-energie}
%\end{figure}

%\begin{figure}
%\vspace{-1cm}
%\begin{center}
%\begin{picture}(150,220)
\put(70,0){\epsfxsize80mm\epsfbox{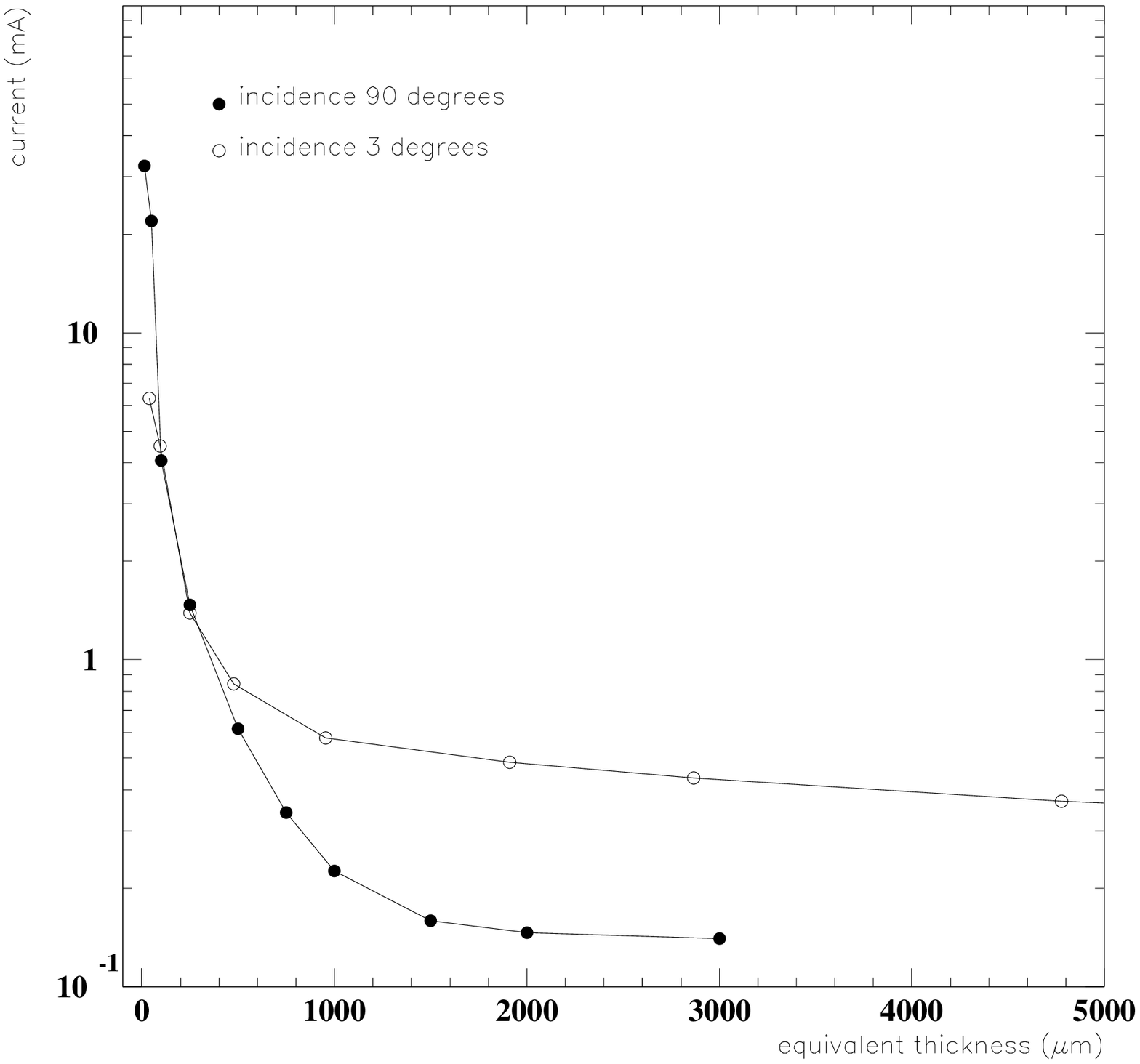}}
\end{picture}
\vspace{-0.5cm}\caption{\protect\footnotesize Power deposit for 3 and 90 degree incidence 
angles and 1 mA current (left) and electron intensity corresponding to a
deposited power of 1 kW  (right), 
as a function of the equivalent thickness of target crossed for a beam energy 
of 10 MeV.}
\label{fig:courantmax}
\end{center}
%\vspace*{-0.5cm} 
\end{figure}

We have computed the energy deposited in targets of various thicknesses with
a current of 1 mA (figure~\ref{fig:courantmax}).
For 50 $\mu$m at 3 degrees or 1mm at 90 degrees, which have the same 
equivalent thickness~\footnote{thickness of target 
material crossed by electrons supposing straight line propagation 
(see figure~\ref{fig:simul-cible}).}, 
the deposited energy is respectively 1750 W and 4500 W. 
This difference is due to the fact that electrons have more possibilities to 
escape the target when the incidence angle is small. The total path length 
of an electron and its secondary electrons inside
the target is on average 5 times longer at 90 degrees than at 3 degrees for the
same equivalent thickness.

The deposited energy increases with equivalent thickness but saturates
earlier with very low incidence angle. This allows to deliver a higher 
intensity for the same illuminated surface.
Figure~\ref{fig:courantmax} shows the electron current intensity which
corresponds to  a deposited energy of 1 kW as a function of the equivalent 
thickness.

\subsection{Production rate}

\begin{figure}
%\vspace{-1cm}
\begin{center}
\begin{picture}(150,220)
\put(-140,0){\epsfxsize80mm\epsfbox{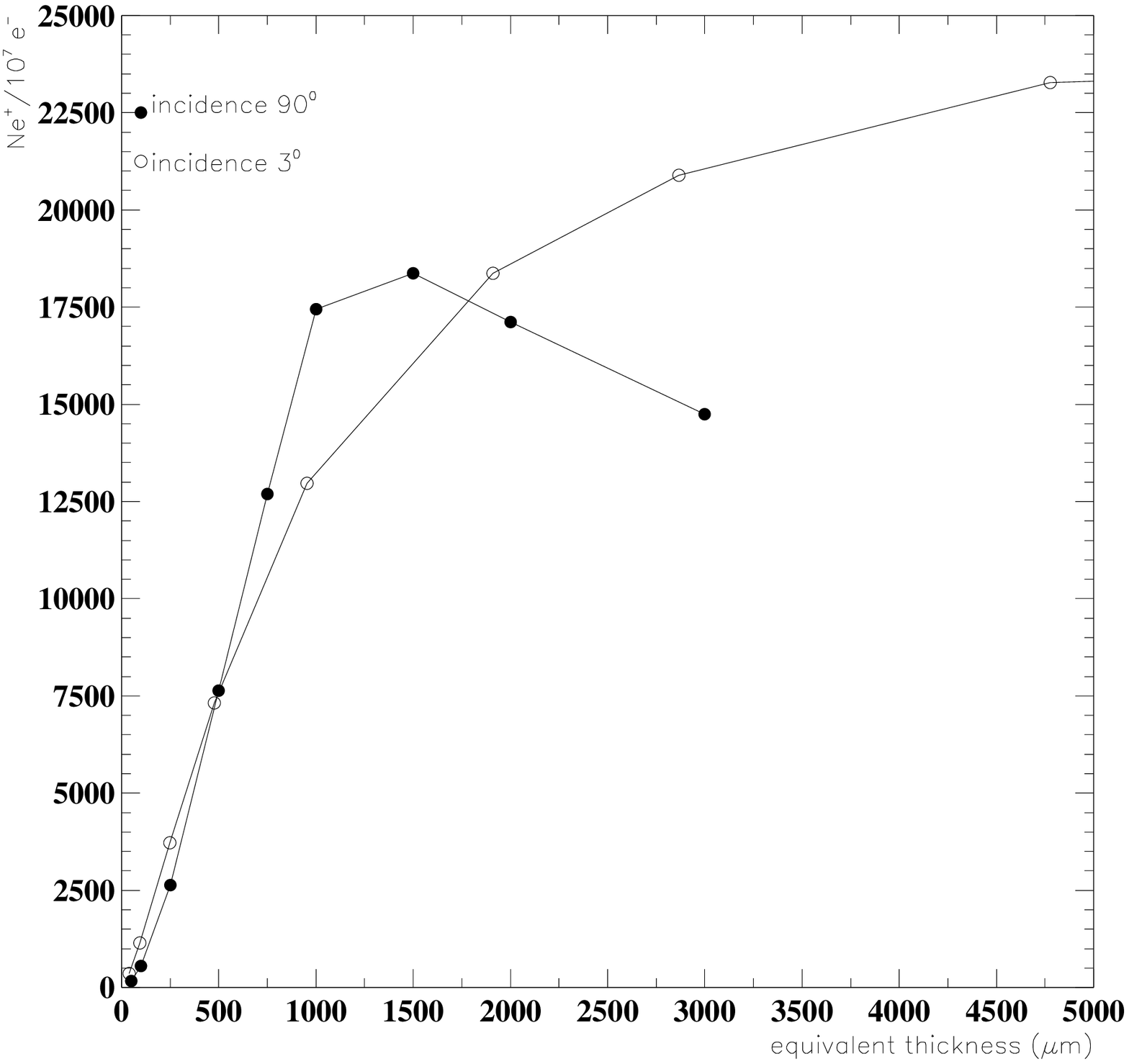}}
%\end{picture}
%\includegraphics*[width=80mm]{prod-10-all.eps}
%\end{center}
%\vspace*{-0.75cm} 
%\caption{\protect\footnotesize Number of positrons produced downstream of the 
%target for $10^7$ electrons generated 
%as a  function of the equivalent target thickness
%crossed for incidence angles of 3 and 90 degrees at a beam energy of 10 MeV.
%}
%\label{fig:prod-all}
%\end{figure}

%\begin{figure}[h]
%\vspace{-1cm}
%\begin{center}
%\begin{picture}(150,110)
\put(70,0){\epsfxsize80mm\epsfbox{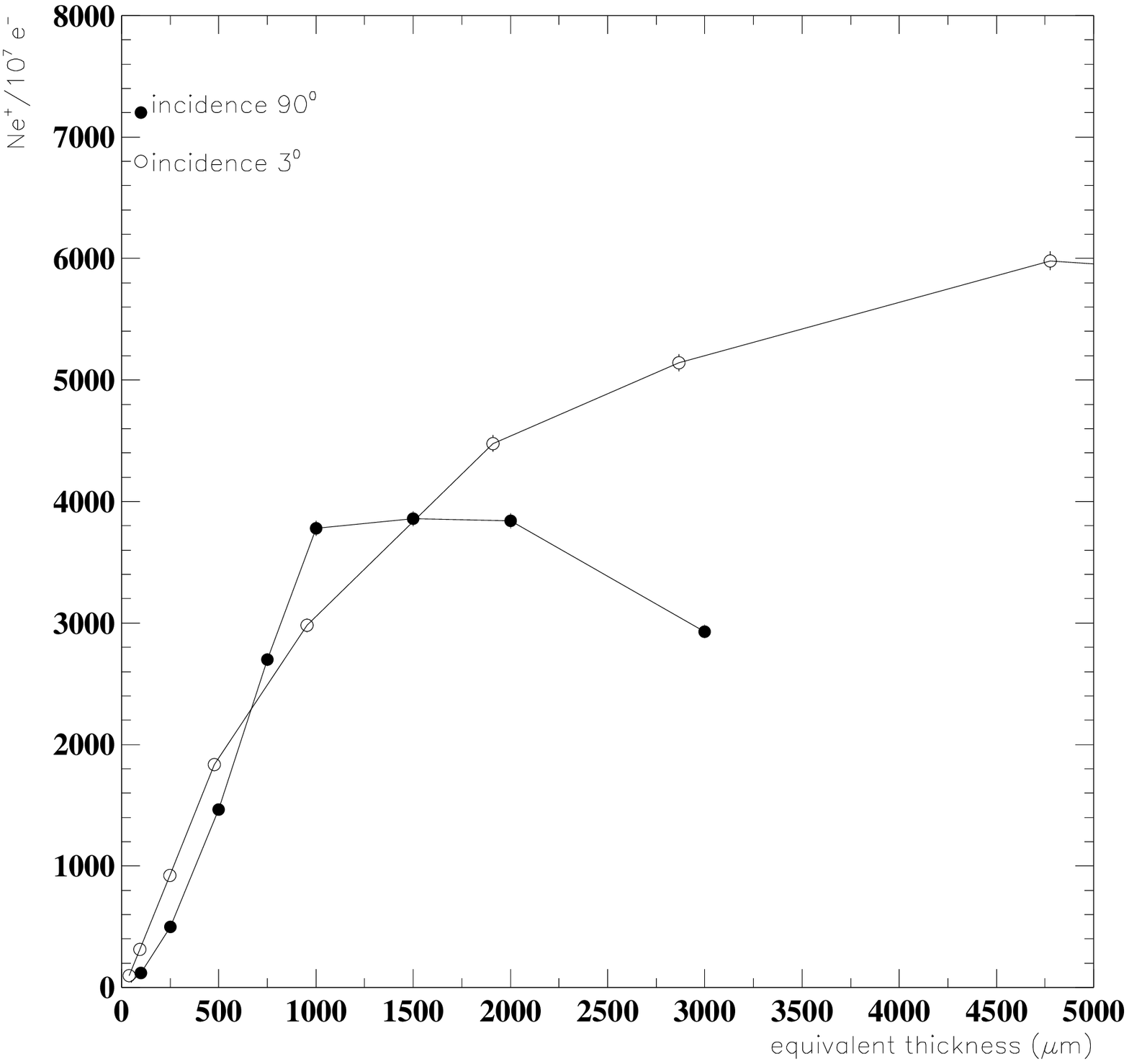}}
\end{picture}
\end{center}
\vspace*{-0.5cm} 
\caption{\protect\footnotesize Number of positrons produced
downstream of the target for $10^7$ electrons generated 
as a  function of the equivalent 
target thickness crossed for incidence angles of 3 and 90 degrees at a beam
energy of 10 MeV. Total (left) and of less than 1 MeV (right).
}
\label{fig:prod}
\end{figure}

The positron rate is given in figure~\ref{fig:prod}.
It is of the order of $10^{13} e^+ s^{-1}$ for 1mA electron current and
equivalent thicknesses between 1 and 2 mm.
These results agree with an independant similar study~\cite{lessner}.
The rate of positrons produced with less than 1 MeV of kinetic energy shown 
in figure~\ref{fig:prod} is about 1/5 of the total rate.

\begin{figure}
%\vspace{-1.5cm}
\begin{center}
\begin{picture}(150,220)
\put(-140,0){\epsfxsize80mm\epsfbox{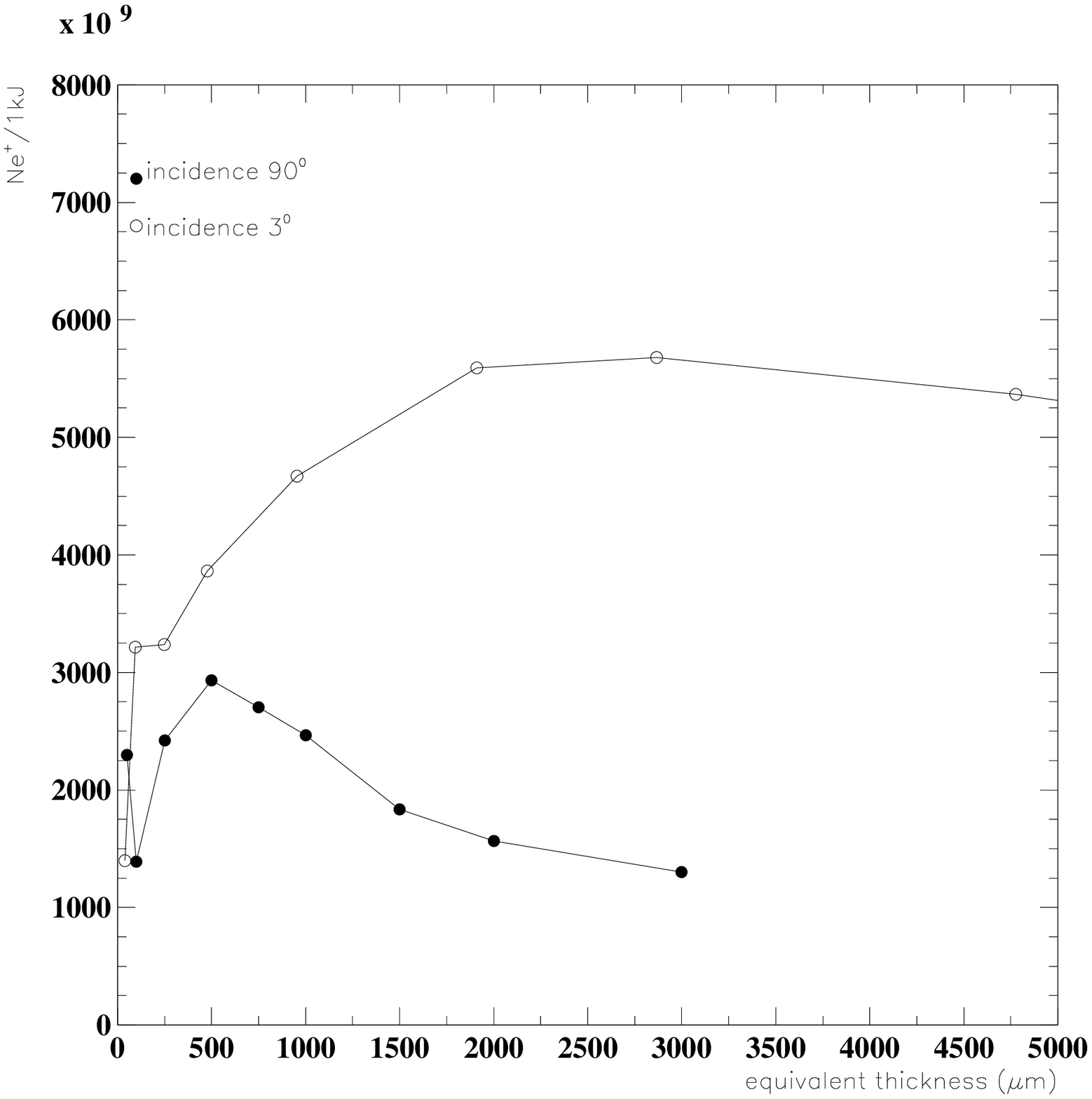}}
%\end{picture}
%\includegraphics*[width=80mm]{ratepos-10-all.eps}
%\end{center}
%\vspace*{-0.75cm} 
%\caption{\protect\footnotesize Produced positron rate downstream of the 
%target for an electron intensity corresponding to a deposited power of 1 kW
%as a function of the equivalent target thickness crossed for 3 and 90 degree 
%incidence angles and a beam energy of 10 MeV.
%}
%\label{fig:ratepos-1kW-all}
%\end{figure}

%\begin{figure}[h]
%\vspace{-1cm}
%\begin{center}
%\begin{picture}(150,110)
\put(70,0){\epsfxsize80mm\epsfbox{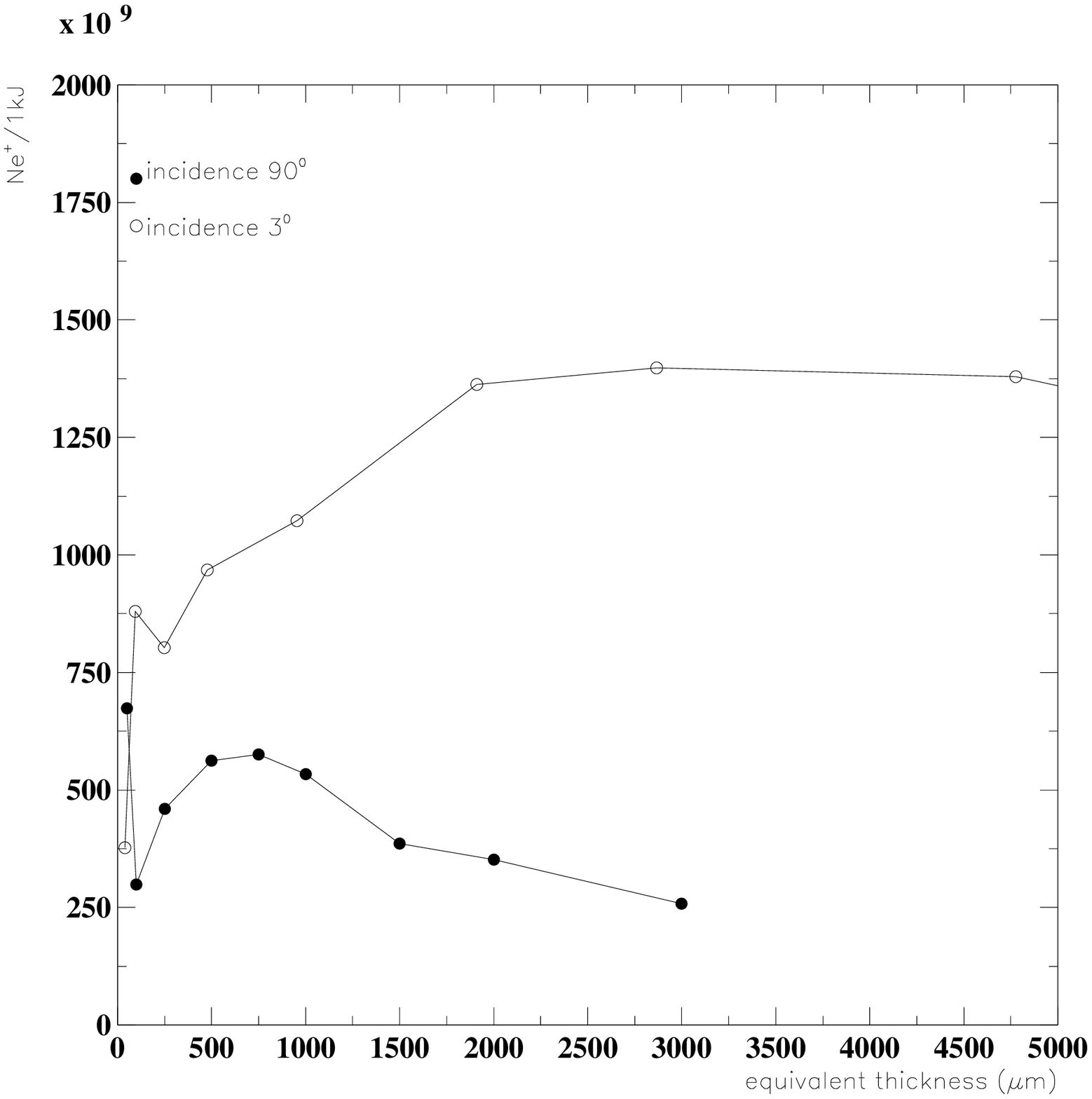}}
\end{picture}
\end{center}
\vspace*{-0.5cm} 
\caption{\protect\footnotesize Produced positron rate 
downstream of the 
target for an electron intensity corresponding to a deposited power of 1 kW
as a function of the equivalent target thickness crossed for 3 and 90 degree 
incidence angles and a beam energy of 10 MeV. Total (left) 
and of less than 1 MeV (right).
}
\label{fig:ratepos-1kW}
\end{figure}

The limit on the deposited power at 1 kW/$\rm{cm^2}$, 
determines the maximum current intensity per $\rm{cm^2}$ 
of target (see figure~\ref{fig:courantmax}). 
The corresponding positron rates are shown in
%figures~\ref{fig:ratepos-1kW-all} and~\ref{fig:ratepos-1kW}.
figure~\ref{fig:ratepos-1kW}.

Let us take the example of a 1 $\rm{cm^2}$ target of 50 micron thickness
at an incidence of 3 degrees or 0.96 mm equivalent thickness. 
The deposited power for a 1 mA electron current is 1.75 kW (resp. 4.5 kW) at 3 
degrees (resp. 90  degrees).
The maximum acceptable current for the limit of 1 kW/$\rm{cm^2}$, as well as 
the corresponding positron rate for the maximum current are given in 
table~\ref{tab:tab1}.

\begin{table}[htbp]
\begin{center}
\caption{\protect\footnotesize E = 10 MeV, target of 
1 $\rm{cm^2} x $50 $\mu$m or 0.96 mm}
\vspace*{0.5cm}
\label{tab:tab1}
\begin{tabular}{|c|c|c|c|}
\hline
E=10MeV & 
$\rm{I_{max}}$ & 
$Ne^{+}$ & $Ne^{+} (<1\rm{MeV})$\\
\hline
3 degrees  & 0.58 mA & 4.6 $10^{12} \rm{s^{-1}}$ & 
1.1 $10^{12} \rm{s^{-1}}$ \\ 
\hline
90 degrees & 0.25 mA & 2.5 $10^{12} \rm{s^{-1}}$ & 
0.55 $10^{12} \rm{s^{-1}}$ \\ 
\hline
\end{tabular}
\end{center}
\end{table}

%Figures~\ref{fig:ratepos-1kW-all} and~\ref{fig:ratepos-1kW} show
Figure~\ref{fig:ratepos-1kW} shows
that these values stay valid within a factor two for equivalent thicknesses 
varying from 500 $\mu$m to 5 mm.

The above quoted production rates were normalized to a target surface of 
1 $\rm{cm^2}$. In the case of the low incidence angle, it is possible to 
increase the beam intensity while keeping a reasonable transverse extension 
of the beam, a key parameter to keep a good efficiency for the collection 
setup described below. 
In order to illuminate a target of size 1 cm x 1 cm at 3
degrees, the beam is a slit of 1 cm x 0.5 mm. 

We will see in the next section on the collection of positrons that it is 
possible to recover a large fraction of the low momentum positrons
with a target size of 2 cm x 2 cm, the beam is then a slit of 2 cm x 1 mm.
The beam current needed to illuminate such a target while keeping the
constraint of 1 kW/$\rm{cm^2}$ is then 2.3 mA which 
produces 4.4 $10^{12} \rm{s^{-1}}$ positrons of less than 1 MeV of kinetic 
energy.

%Active cooling by water flow is possible but with added complexity.
Moving a long strip of 50 $\mu m$ thick tungsten in the target plane is a 
preferred solution to increase the beam current above 2.3 mA.

\subsection{Collection}

\begin{figure}[htbp]
%\vspace{-1cm}
\begin{center}
\begin{picture}(150,220)
\put(-140,0){\epsfxsize80mm\epsfbox{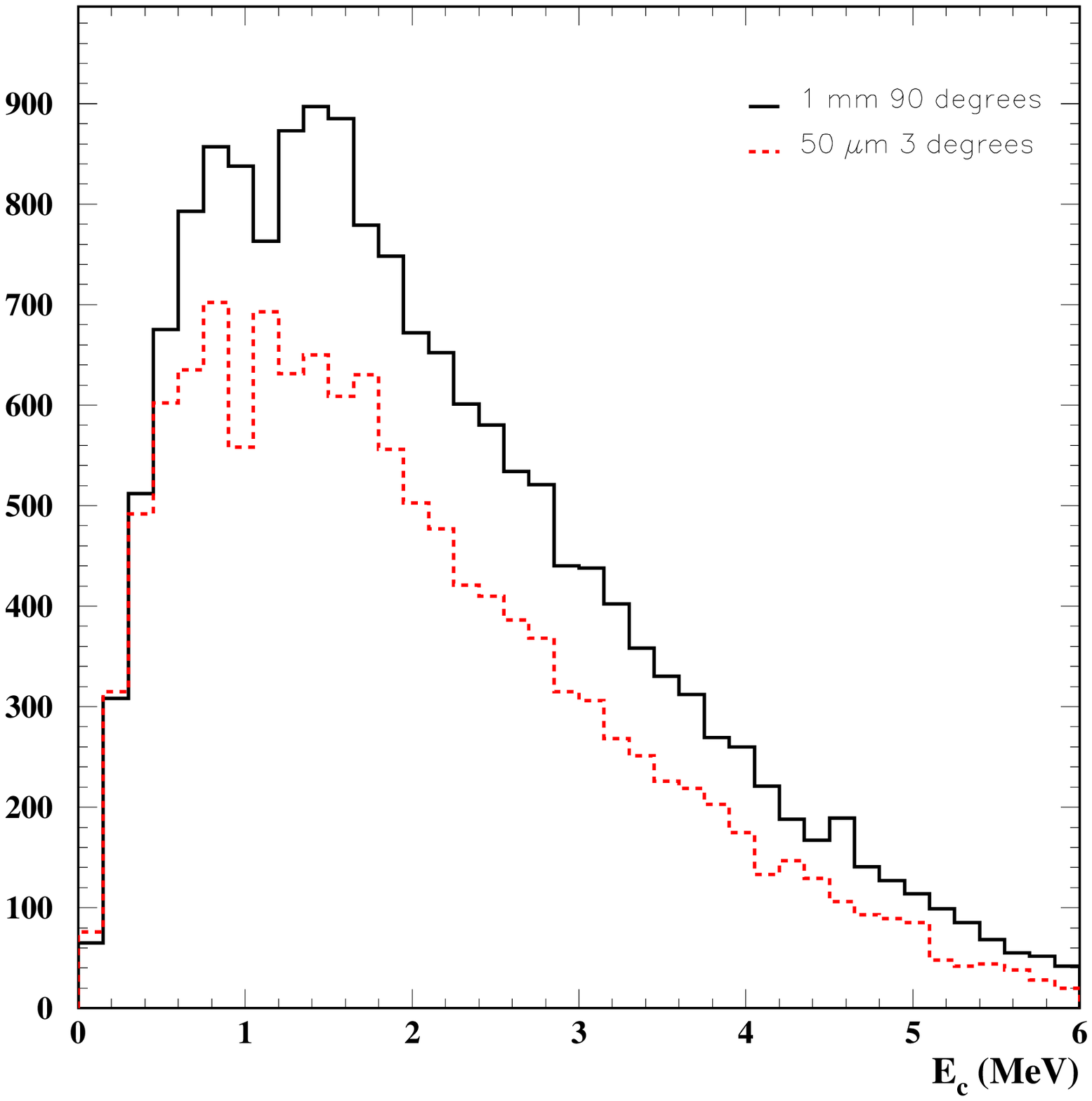}}
\put(70,0){\epsfxsize80mm\epsfbox{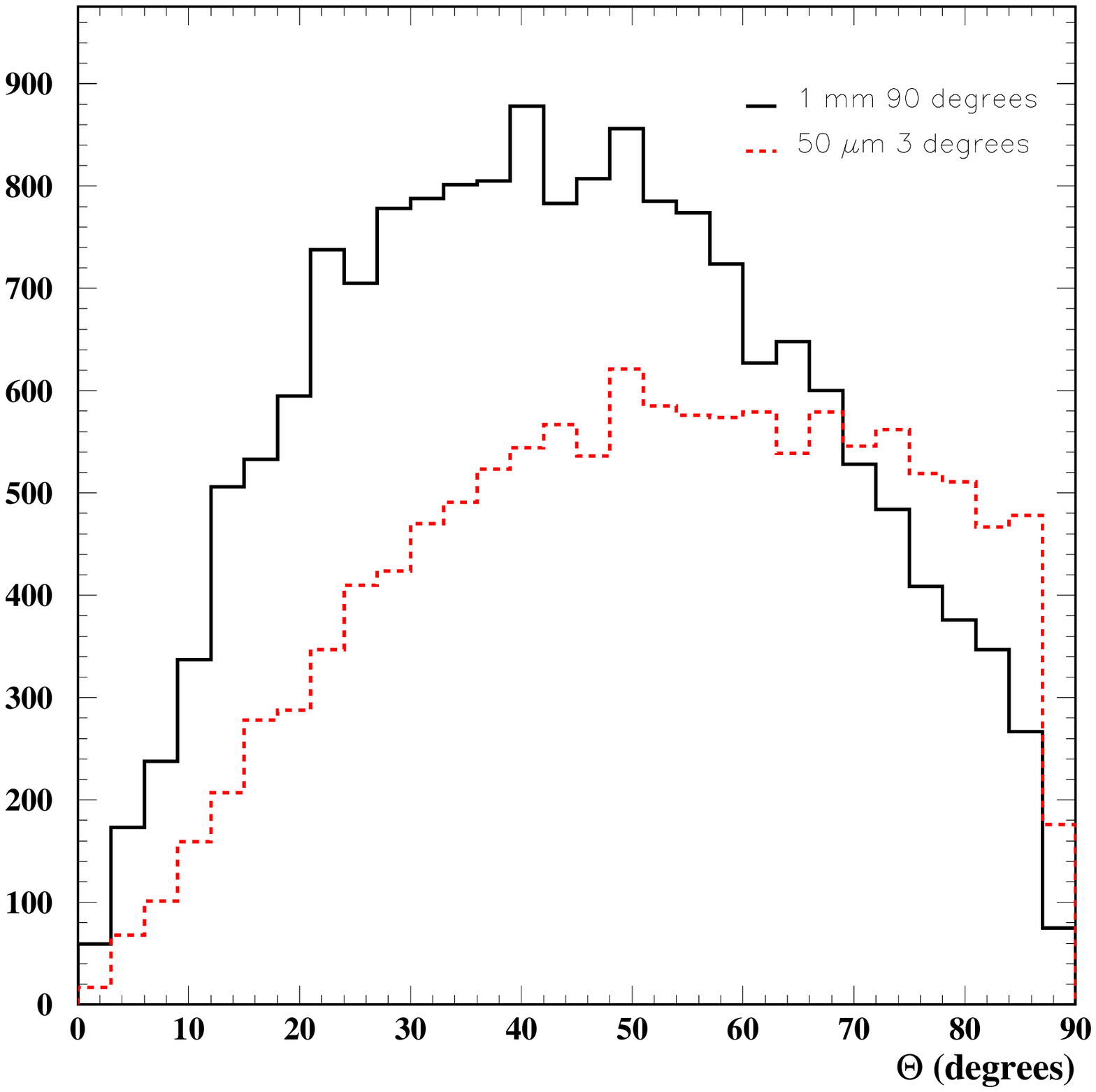}}
\end{picture}
%\includegraphics*[width=80mm]{ekin-10.eps}
%\vspace*{-0.5cm} 
%\includegraphics*[width=80mm]{theta-10.eps}
\end{center}
\vspace*{-0.5cm} 
\caption{\protect\footnotesize Spectra of the kinetic energy and 
production angle of positrons downstream of the target for
$10^7$ \'{e}lectrons
with 3 and 90 degree incidence angles at a beam energy of 10 MeV.
Distributions are normalized to the number of electrons generated.}
\label{fig:ekinthet_10}
\end{figure}

Figure~\ref{fig:ekinthet_10} shows the energy distribution of positrons 
downstream of the target and their angular distribution. 
The main feature is that the average exit angle with respect to the beam
is large, of the order of 50 degrees and even larger for the lowest positron 
energies.

It is thus necessary to develop a positron collector in order to transport
them efficiently at the trap entrance. In order to take advantage of this
wide angle of production, a system of coils 
producing diverging magnetic field lines at the location of the target is used.
The magnetic lines are collected by a large diameter coil to form a magnetic 
bulb.

\subsection{Description of the setup}
The $x$ axis is the axis of the apparatus. The target is a thin rectangular 
tungsten plate of dimensions 2 cm x 2 cm x 50 $\mu$m. There are two possible 
ways to place the beam: 
\begin{itemize}
\item setup 1: 
the beam axis coincides with the $x$ axis, in which case the target plane
makes a 3 degree angle with the $x$ axis, 
\item setup 2: 
the beam axis $x'$ makes an angle of 3 degrees with axis $x$, and the
target plane contains the $x$ axis.
\end{itemize}

The beam has the shape of a rectangular slit of 2 cm x 1 mm.

In the second configuration, a large part of the beam which traverses the 
target without much deflection separates from the $x$ axis.

\begin{figure}[htbp]
%\vspace{-1cm}
\begin{center}
%\begin{picture}(150,220)
%\put(5,5){\epsfig{totalheight=12cm, angle=90, file=setupX.eps}}
%\end{picture}
\includegraphics*[width=120mm, angle=0]{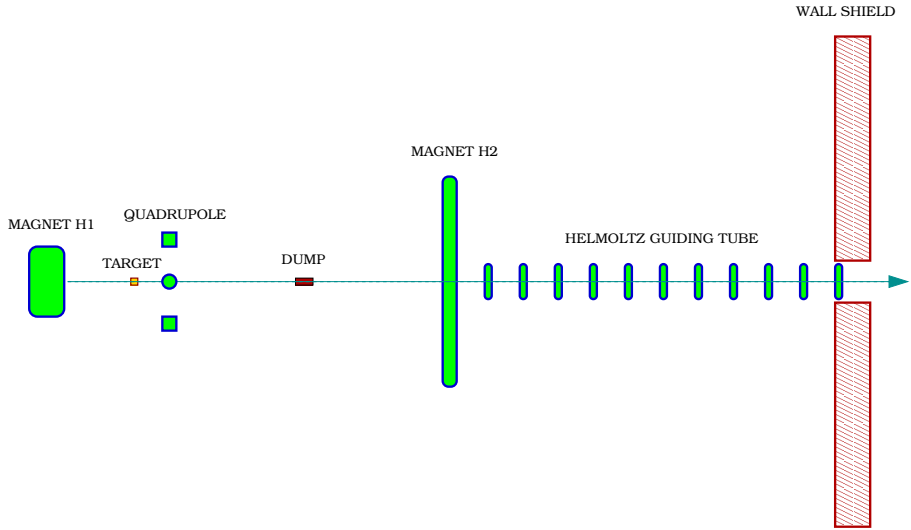}
\end{center}
%\vspace*{-0.5cm} 
\caption{\protect\footnotesize Layout of the proposed magnetic collection
system.}
\label{fig:trieuse}
\end{figure}

\begin{figure}[htbp]
%\vspace{-1cm}
\begin{center}
%\begin{picture}(150,220)
%\put(5,5){\epsfig{totalheight=15cm, angle=90, file=posit-9trk.eps}}
%\end{picture}
\includegraphics*[width=120mm, angle=0]{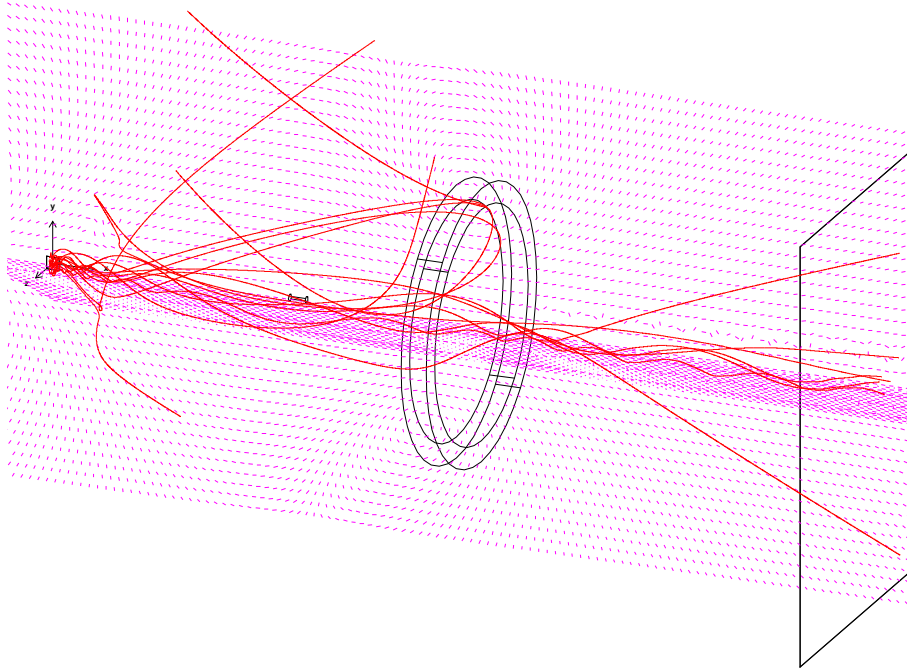}
\end{center}
%\vspace*{-0.5cm} 
\caption{\protect\footnotesize Positron tracks in the collection system: 
below 1 MeV the particles are guided to the exit tube.}
\label{fig:collect}
\end{figure}

\begin{table}[htbp]
\begin{center}
\caption{\protect\footnotesize Layout scales}
\vspace*{0.5cm}
\label{tab:layout}
\begin{tabular}{|c|c|}
\hline
item & x (cm) \\
\hline
H1 superconducting coil & -20 \\
target & 0 \\
quad & 11 \\
tungsten dump & 51 \\
H2 large collecting coil & 90 \\
first exit tube coil& 100 \\
\hline
\end{tabular}
\end{center}
\end{table}

The collection system consists of two main coils of axis $x$ 
(figure~\ref{fig:trieuse}). 
The first coil, 
$H_1$, has a mean radius of 10 cm, a width along $x$ of 5cm and a 5 Tesla
field at its center. The target center is placed 20 cm downstream the center 
of $H_1$.

The second coil $H_2$ has a 30 cm radius, 20 kA.turns, producing in its center
a 420 Gauss field in the same direction as the field at the center of $H_1$. 
It is placed 90 cm downstream of the center of the target.

A ``recovery'' tube consists of a series of flat coils
of axis $x$ in Helmholtz configuration one with respect to the next, each with
a 10 cm diameter and 2 kA.turns. The first of these small coils is placed 
10 cm downstream of $H_2$. The following ones are placed 7 cm from each other
constituting a kind of open solenoid. 

The beam goes through $H_1$ and hits the target. The particles exiting the 
target go then through $H_2$ and then through the recovery tube.

A small quadrupole of 10 cm internal radius is placed 11 cm downstream of 
the target. 
Its four coils have each a 3 cm radius and 2kA.turns, or a  250 Gauss field 
at their center. 
Two of its coils are vertical and the other two are horizontal.
It pulls the positrons nearer to the $x$ axis.

The field on the $x$ axis at the center of the target is of 0.46 Tesla. The
radial component at a radius of 1 cm from this axis is 0.28 Tesla.

A dump made of tungsten of 1.6 cm diameter and 5 cm length is placed on the 
$x$ axis at 51 cm downstream of the target.

In order to illuminate the target with a simple slit shaped electron beam,
it is enough for setup 1 to include a tilt angle of the target
with respect to the vertical
in order to correct for the deviation from coil $H_1$. For setup 2 a
beam optics has been designed using octupoles following a method developed by
F. Meot~\cite{meot}.

\subsection{Collection efficiency}

Let us define the collection efficiency $\epsilon_5$ (resp.$\epsilon_2$)
as the fraction of positrons of a given energy range reaching a plane 
perpendicular to the $x$ axis and located inside the recovery tube at a 
distance of 2 m from the target, so 1.1 m after $H_2$, 
and whose intersection point with this plane is 
inside a circle of radius 5 cm (resp. 2 cm).

\begin{figure}[htbp]
\vspace{-5mm}
\begin{center}
\begin{picture}(150,220)
\put(-140,0){\epsfxsize80mm\epsfbox{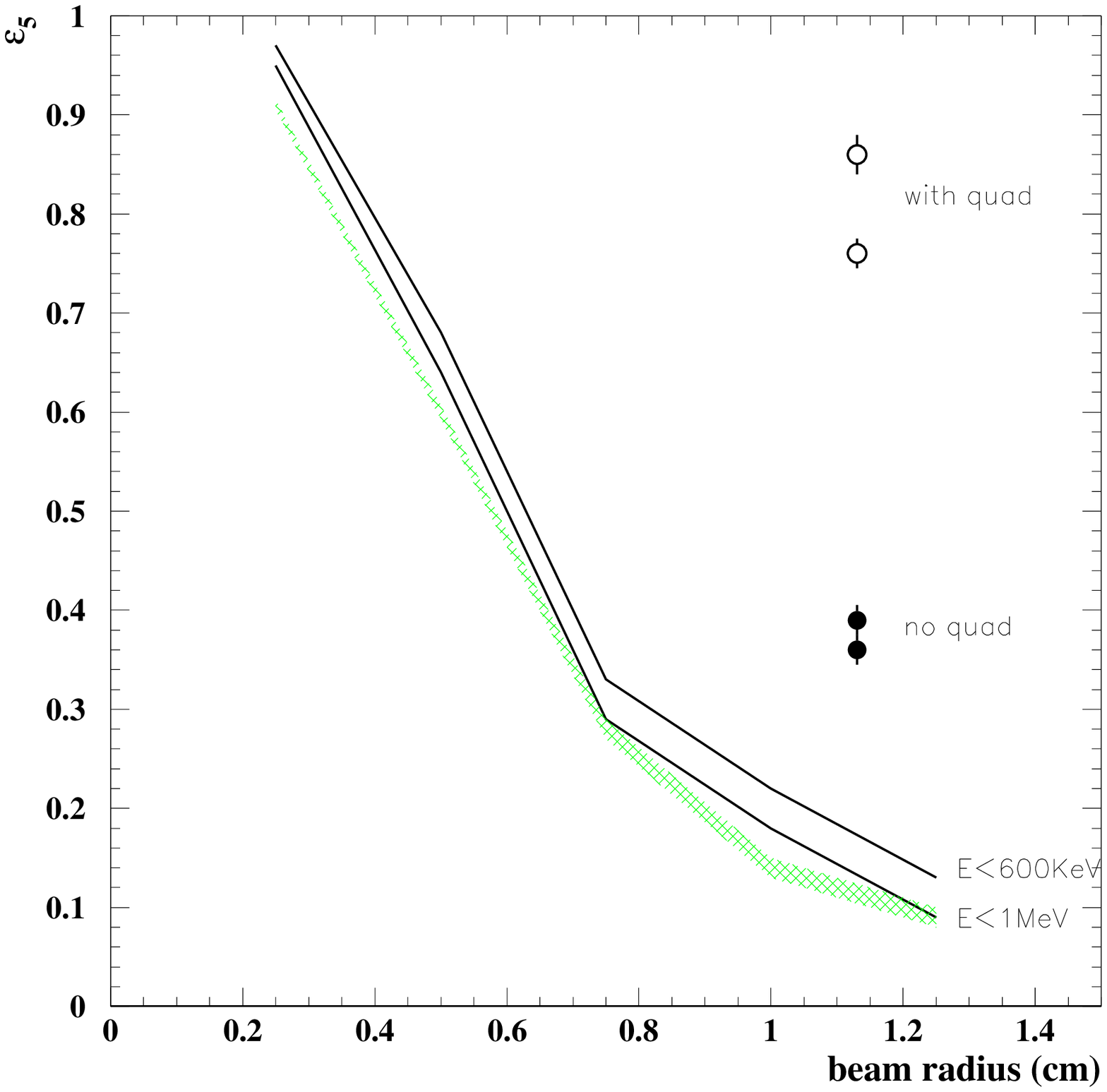}}
\put(70,0){\epsfxsize80mm\epsfbox{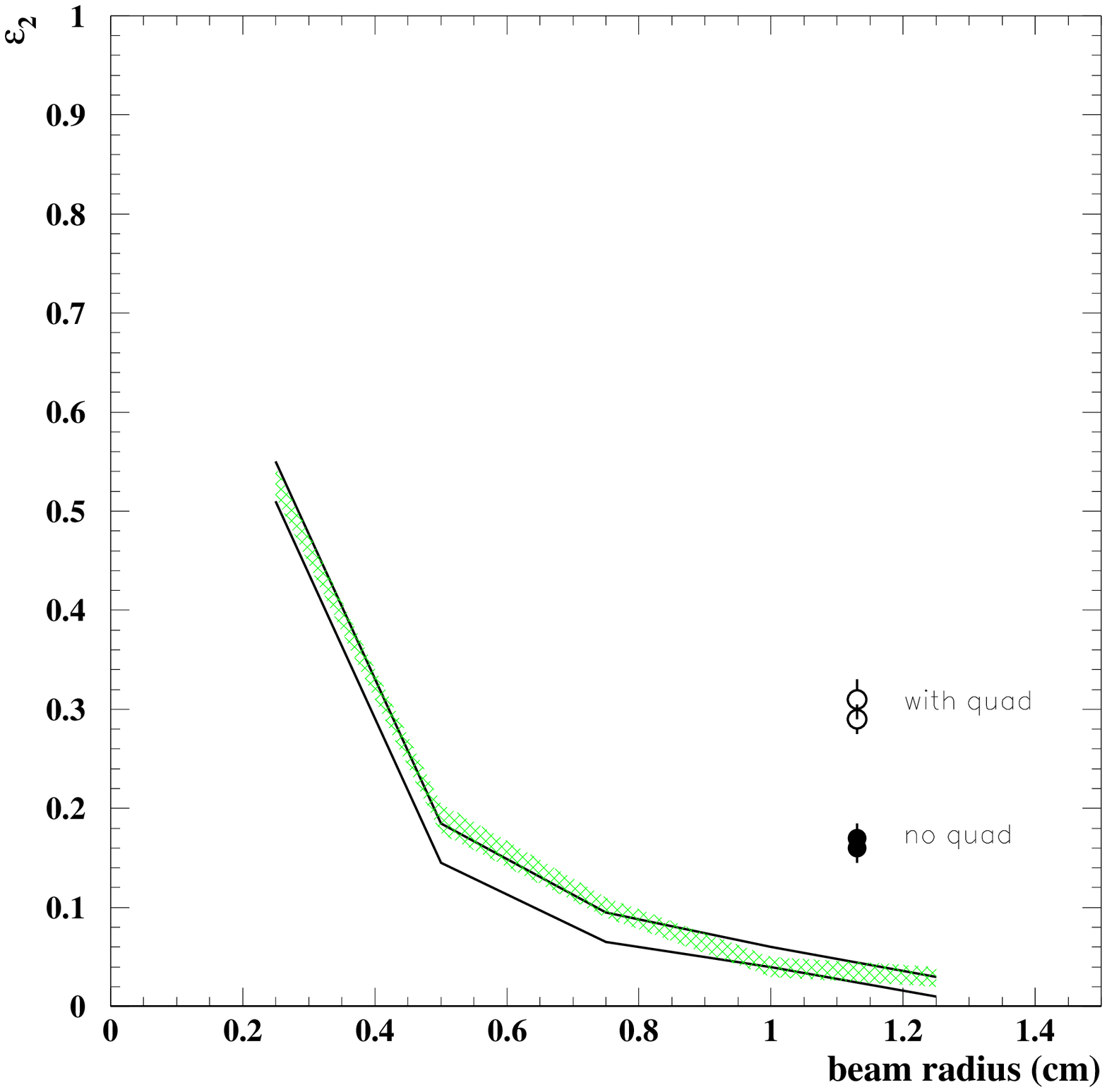}}
\end{picture}
%\includegraphics*[width=80mm]{eff_rad5.eps}
%\vspace*{-0.5cm} 
%\includegraphics*[width=80mm]{eff_rad2.eps}
\end{center}
\vspace*{-0.75cm} 
\caption{\protect\footnotesize 
Efficiency as a function of the radius of a disk shaped beam with uniform 
distribution for kinetic 
energies below 600 KeV. The hatched band corresponds to kinetic energies below
1 MeV. The points correspond to a 2 cm x 1 mm rectangular slit shaped beam 
parallel to the axis of the apparatus with or without correcting quadrupole.
Left (resp. right) curves are for $\epsilon_5$ (resp. $\epsilon_2$)}
\label{fig:eff_rad}
\end{figure}

This efficiency depends on the transverse spread, with respect to the $x$ axis,
of the positron ``beam spot''. 
Figure~\ref{fig:eff_rad} shows the efficiency variation
with the radius of a hypothetical positron source located at the target 
position. For this calculation, we have generated a uniform spatial 
distribution of the emitted positrons inside a disk perpendicular to the $x$ 
axis. 
The angular and energy spectrum corresponding to that of positrons exiting the
tungsten target was kept. 
The efficiency decreases dramatically when the radius of this disk is 
increased.

On the same figure is also shown the efficiency obtained
with a rectangular slit of 2 cm x 1 mm, the $x$ axis being contained in the 
target plane, with or without correcting quadrupole.
The same efficiency would be obtained with a disk shape if it had a radius of
about 0.4 cm, which could only be obtained with a reduction of the surface of
target illuminated by a factor of 0.13 with respect to the 4 $\rm{cm^2}$ plate,
and thus a reduction of about an order of magnitude in the rates.

Values for $\epsilon$ and the final rates of positrons at 2 m from the target
are given in table~\ref{tab:tocoll}.

\begin{table}[htbp]
\begin{center}
\caption{\protect\footnotesize Collection efficiency and collection rates in 
units of $10^{12} s^{-1}$ }
\vspace*{0.5cm}
\label{tab:tocoll}
\begin{tabular}{|c|c|c|c|c|}
\hline
    & 
$\epsilon_{5}$ & 
$e^+_5$ rate   & $\epsilon_{2}$ & 
$e^+_2$ rate \\
\hline
Ec $>$ 0  & 20\%   & 4.2  & 8\%   & 1.6   \\ 
\hline
Ec $<$ 1 MeV  & 52\%   & 2.7  & 20\%   & 1.1   \\ 
\hline
Ec $<$ 600 KeV  & 60\%   & 1.3   & 30\%   & 0.6    \\ 
\hline
\end{tabular}
\end{center}
\end{table}

\subsection{Flux of electrons and photons}

The power coming from this device in the space 
downstream of the coils affects the design of the room shielding, and
the coupling to the trap which has a first stage at very low temperature.

\begin{figure}
%\vspace{5cm}
\begin{center}
\begin{picture}(150,220)
\put(-140,0){\epsfxsize80mm\epsfbox{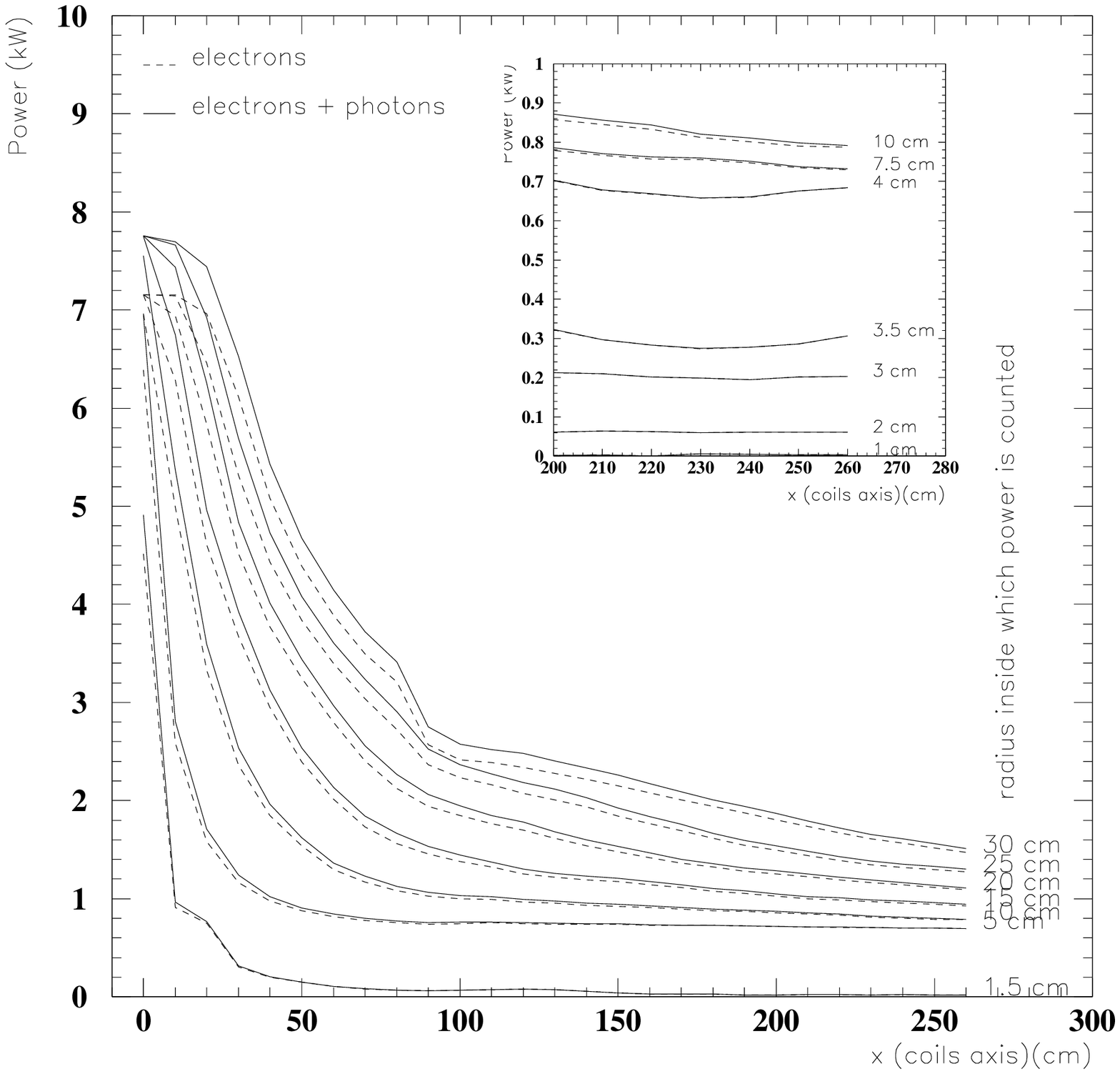}}
%\end{picture}
%\includegraphics*[width=80mm, angle=0]{flux-stret.eps}
%\end{center}
%\vspace*{-1cm} 
%\caption{\protect\footnotesize {\bf Setup 1}. 
%Flux of electrons and photons (full line)
%and electrons (dashed line) within circular sectors around the setup axis 
%as a function of the distance to the target 
%for several radii of the circular sectors. The total beam power is 10 kW.}
%\label{fig:flux-stret}
%\end{figure}

%\begin{figure}[htbp]
%\vspace{2cm}
%\begin{center}
%\begin{picture}(150,130)
\put(70,0){\epsfxsize80mm\epsfbox{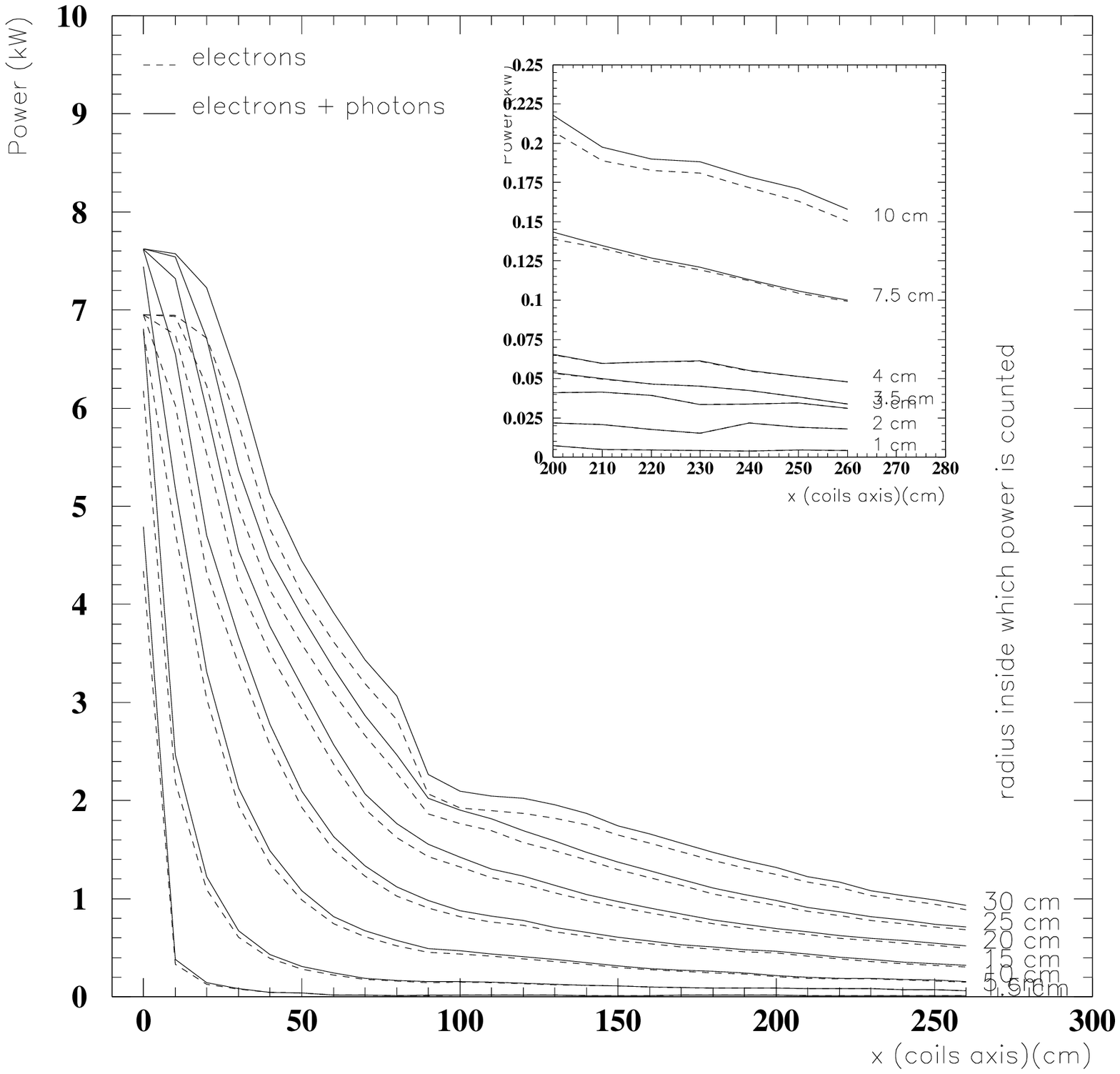}}
\end{picture}
\end{center}
\vspace*{-0.5cm} 
\caption{\protect\footnotesize  
Flux of electrons and photons (full line)
and electrons (dashed line) within circular sectors around the setup axis 
as a function of the distance to the target 
for several radii of the circular sectors. The total beam power is 10 kW.
Setup 1 (left) and Steup 2 (right).}
\label{fig:flux-stret}
\end{figure}

%Figures~\ref{fig:flux-stret} and~\ref{fig:flux-800ka}
Figure~\ref{fig:flux-stret} 
shows for setup 1 and 2 the flux of electrons and photons which cross 
planes perpendicular to the $x$ axis, and whose intersection with these planes
lie within circles of various radii, as a function of
the distance between these planes and the target, for a 1 mA electron beam.
The beam is supposed to illuminate the totality of the target surface.
At target exit, 7.8 kW are collected. The power deposited in the target is 
1.8 kW. The missing 400 W are back scattered.

The power reaching a plane located at 2.5 m from the target is given in detail
in table~\ref{tab:power} for an electron current of 2.3 mA.

\begin{table}[htbp]
\begin{center}
\caption{\protect\footnotesize Power at 2.5 m from target as a function of 
radius, for 2.3 mA.}
\vspace*{0.5cm}
\label{tab:power}
\begin{tabular}{|c|c|c|c|c|}
\hline
  & 
R = 1 cm & R = 2 cm & R = 3 cm & R = 4 cm  \\
\hline
setup 1 & 5 W   & 140 W & 450 W & 1.5 kW  \\ 
\hline
setup 2  & 10 W   & 45 W & 80 W & 110 W \\ 
\hline
\end{tabular}
\end{center}
\end{table}

\subsection{Summary}

A system of production and collection of MeV positrons has been presented.
This setup uses pair creation from electrons hitting a 
thin tungsten target at a 3 degree incidence angle.
The beam comes from a 10 MeV/2.3 mA electron accelerator running
in continuous mode.
The setup allows to collect more than 5 $10^{11} s^{-1}$ positrons of less 
than 600 keV in a 2 cm radius aperture at 2 m from the target.
The flux of electrons and photons reaching this aperture is of the order of 
50 W.

The system is designed to adapt to a Greaves-Surko trap.

The source itself is made of two sections: \\
- the 10 MeV accelerator with a dump\\
- the target and collector system

The best electron source for this application is a Rhodotron, which is
commercialized by the IBA~\cite{IBA} firm. 
This compact machine (about 3 m in diameter) is operated continuously all year 
long and can reach beam intensities up to 100 mA for food disinfection.
Several models are available according to the desired beam intensity. 
A 10 MeV/2mA model would use 250 kW of 
power.

{\bf Acknowledgements}\\
We wish to express our sincere thanks to all the people from different fields with 
whom we had fruitful discussions in order to design this project:
B. Aune,
G. Baldacchino,
J.L. Borne,
P. Debu,
S. Ecklund,
R. Greaves,
M. Jablonka,
P. Lecoeur,
B. Mansouli\'{e},
F. M\'{e}ot,
J.C. Mialocq,
A.P. Mills,
A. N'Guyen,
J. Sheppard,
M. Spiro,
C. Surko,
G. Vigneron,
and M. Woods.

This technique is subject to a US patent (Serial No: 60/470,883).


\bigskip


\begin{thebibliography}{00}

\bibitem{traps}T.J. Murphy and C.M. Surko, Phys. Rev. {\bf A 46} (1992) 5696;
C.M. Surko, S.J. Gilbert and R.G. Greaves, Non-Neutral Plasma Phys. III, 
edited by J.J. Bollinger, R.L. Spencer and R.C. Davidson, 
(American Institute of Physics, New York), 3-12 (1999);\\
http://physics.ucsd.edu/research/surkogroup/positron/buffergas.html

\bibitem{plasma2}R.G. Greaves and C.M. Surko, Phys. Plasmas {\bf 4} (1997) 
1528.\\ S.J. Gilbert et al., Phys. plasmas {\bf 8} (2001) 4982.
 
\bibitem{atom2}M. Charlton, and J.W. Humbertson, Positron Physics, 
(Cambridge Univ. Press, 2001); C.M. Surko, New Directions in 
Antimatter Chemistry and Physics, Surko and Gianturco eds, 2001, 
Kluwer Academic Publishers.

\bibitem{exp-ad}M. Amoretti et al., Nature, {\bf 419} (2002) 456;
G. Gabrielse et al., Phys. Lett. {\bf B 548} (2002) 140;
G. Gabrielse et al., Phys. Rev. Lett. {\bf 89}, 233401 (2002);
G. Gabrielse et al., Phys. Rev. Lett. {\bf 89}, 213401 (2002).

\bibitem{astro2}K. Iwata, R.G. Greaves and C.M. Surko, Can. J. Phys. {\bf 51}
(1996) 407.

\bibitem{materials}P.J. Schultz and K.G. Lynn, Rev. Mod. Phys. {\bf 60} (1988)
701.

\bibitem{rotatfield}R.G. Greaves and C.M. Surko, Phys. Plasmas {\bf 8} (2001) 
1879.
\bibitem{our-antiH}Paper in preparation.

\bibitem{mills-imaging}A.P. Mills and P.M. Platzman, New Directions in 
Antimatter Chemistry and Physics, p. 115, Surko and Gianturco eds, 2001, 
Kluwer Academic Publishers.

\bibitem{mills-laser}A.P. Mills, Nucl. Inst. Meth. {\bf B 192} (2002) 107.

\bibitem{Liang}E.P. Liang, C.D. Dermer, Opt. Commun. {\bf 65} (1988) 419.

\bibitem{PALS} Radiation Physics and Chemistry {\bf 68} (2003) 329-680.

\bibitem{frm2} C. Hugenschmidt, Nucl. Inst. Meth. {\bf B 192} (2002) 97.\\
FRM-II web site: http://www.frm2.tu-muenchen.de/positron/index.html

\bibitem{surko-trap} R. Greaves et C.M. Surko, Nucl. Inst. Meth. {\bf B 192} 
(2002) 90.

\bibitem{mills-alsio2}
N. Suzuki et al., 
%"Study of Silica Aerogel Grain Surfaces by 
%Using a Positron Age-Momentum Correlation Technique"
Appl. Phys. {\bf A 74} (2002) 791-795. 

\bibitem{goodfellow}Goodfellow SARL, 229, rue Solf\'{e}rino, Lille, F-59000, 
France.

\bibitem{langmuir} Handbook of Chemistry and Physics, 10-305, 
D.R. Lide ed., 73rd edition, 1992, CRC Press.

\bibitem{IBA}IBA (Ion Beam Applications),
Chemin du Cyclotron, 3 - 1348 Louvain-la-Neuve, Belgium, 
http://www.iba-tg.com/root\_hq/index.htm

\bibitem{geant}GEANT 3.21, CERN Library.

\bibitem{lessner}E. Lessner, Proc. 1999 Particle Accelerator Conference, 
New York, p1967.

\bibitem{meot}F. Meot and T. Daniel, Nucl. Inst. Meth. {\bf A 379} (1996) 196.
 

\end{thebibliography}
\end{document}